\documentclass{webofc}
\usepackage[varg]{txfonts}   
\usepackage{listings}
\usepackage{hyperref}
\begin{document}
\title{Four years' interferometric observations of Galactic binary Cepheids}
%
%

\author{
	\firstname{A.} \lastname{Gallenne}\inst{1},
	\firstname{A.}
	\lastname{M\'erand}\inst{1}, 
	\firstname{P.}
	\lastname{Kervella}\inst{2},
	\firstname{N.}
	\lastname{R. Evans}\inst{3},
	\firstname{C.}
	\lastname{Proffitt}\inst{4},
	\firstname{G.}
	\lastname{Pietrzy\'nski}\inst{5}
	 \and
    \firstname{W.}
    \lastname{Gieren}\inst{6}
}

\institute{European Southern Observatory, Alonso de C\'ordova 3107, Casilla 19001, Santiago 19, Chile
\and
			 LESIA, Obs. de Paris, CNRS UMR 8109, UPMC,Univ. Paris 7, 5 Pl. Jules Janssen, 92195 Meudon, France
\and
			Smithsonian Astrophysical Observatory, MS 4, 60 Garden Street, Cambridge, MA 02138, USA
\and
			Space Telescope Science Institute, 3700 San Martin Drive, Baltimore, MD 21218, USA
\and
			Nicolaus Copernicus Astro. Centre, Polish Academy of Sci., Bartycka 18, PL-00-716 Warszawa, Poland
\and
           Universidad de Concepci\'{o}n, Departamento de Astronom\'{i}a, Casilla 160-C, Concepci\'{o}n, Chile
          }

\abstract{%
We give an update on our long-term program of Galactic Cepheids started in 2012, whose goal is to measure the visual orbits of Cepheid companions. Using the VLTI/PIONIER and CHARA/MIRC instruments, we have now detected several companions, and we already have a good orbital coverage for several of them. By combining interferometry and radial velocities, we can now derive all the orbital elements of the systems, and we will be soon able to estimate the Cepheid masses.

}
\maketitle

\section{Introduction}\label{sec:intro}

Cepheids are powerful astrophysical laboratories providing fundamental clues for studying the pulsation and evolution of intermediate-mass stars. However, one of the most critical parameters, the mass, is a long-standing problem because of the 10-20\,\% difference between masses predicted from stellar evolution and pulsation models. Cepheids in binary systems are the only tool to constrain models and make progress on this mass discrepancy. Studying Cepheid's companions can also provide insight on the impact of binarity on the calibration of the period-luminosity relation from the Baade-Wesselink technique and the IR surface brightness method. As many of the companions are blue main-sequence stars, the flux contribution in the near-IR might is often negligible ($<$\,1-2\%), but can be as large as 10\,\% in $V$.

Most of the companions are too close to the Cepheid ($<$\,40 mas) to be resolved with single-dish 8-meter class telescopes, and the high-contrast between the companions and the Cepheids makes the detection even more difficult. But long-baseline interferometry (LBI) is able to reach high-spatial resolution and high-dynamic range.

\section{Update and results}\label{sec:status}

\begin{figure*}
	\centering
	\resizebox{\hsize}{!}{\includegraphics{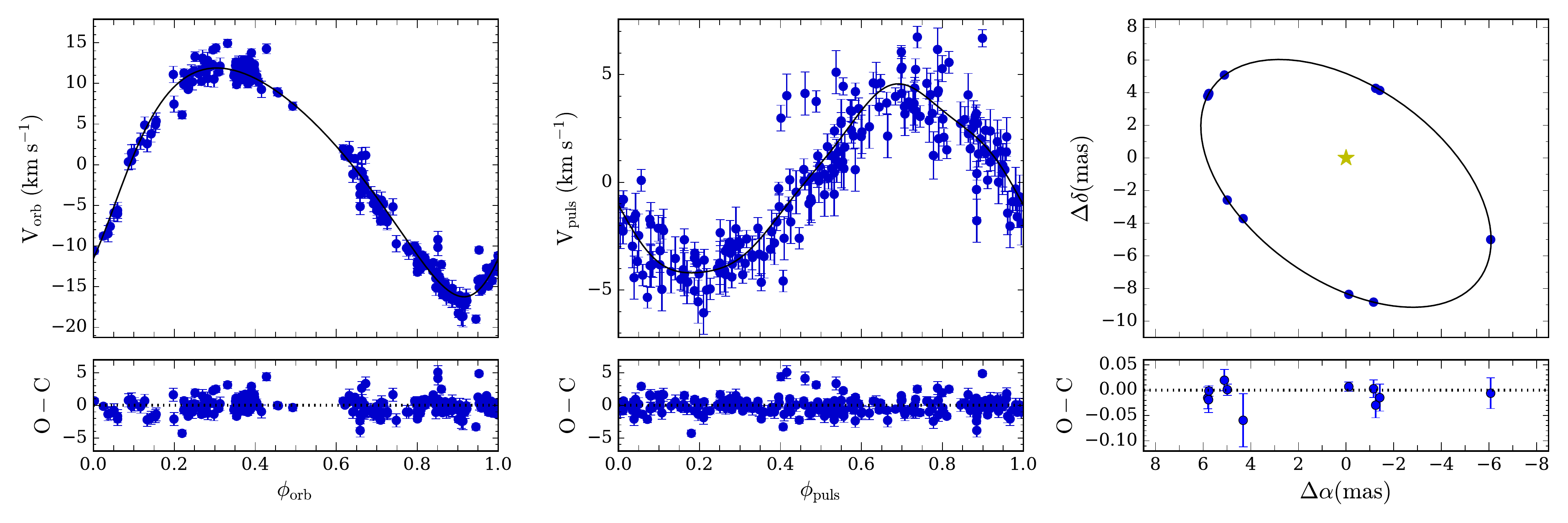}}
	\caption{Combined fit of single-line velocities with astrometry for the Cepheid V1334~Cyg.}
	\label{fig:v1334}       
\end{figure*}

Since 2012 we have observed some binary Cepheids in both the northern and southern hemispheres with the interferometric instruments CHARA/MIRC and VLTI/PIONIER. They are bright Cepheids ($H < 7$\,mag), and single-line spectroscopic binaries with long orbital periods ($\sim 1-40$\,yr). Because of the brightness of the Cepheid, it is difficult to spatially/spectrally detect companions using ground-based telescopes. However, the companion causes the Cepheid to wobble (as the two components orbit their center of mass), which is more easily detectable from radial velocities. 

Using LBI, we detected companions separated by less than 50\,mas from the Cepheids with contrast as high as 6.5\,mag in $H$. For this purpose, we have created a dedicated tool, named \texttt{CANDID}, which searches for high-contrast companions from interferometric data \citep{Gallenne_2015_07_0}. It allows a systematic search on a $N\times N$ grid, whose minimum needed resolution is estimated a posteriori. The tool delivers, among other things, the flux ratio $f$, the astrometric separation $(\alpha,  \delta)$, and the (non-)detection level of the companion based on $\chi^2$ statistics. So far, we have detections for six binary Cepheids, with projected separations ranging from 1.5 to 40\,mas and flux ratios (in $H$) from 0.8 to 4\,\%. The astrometric positions can then be combined with the radial velocities of the primary, as shown in Fig.\ref{fig:v1334} for V1334~Cyg, to provide the full set of orbital elements, including $a, i$ and $\Omega$ previously unknown \citep{Gallenne_2013_04_0,Gallenne_2014_01_0}. However, the distance and masses are still degenerate, and unfortunately no accurate parallaxes exist for these binary Cepheids.

From space, ultraviolet observations from HST/STIS will provide the spectra of the companions. However, broad features (some of them are fast rotators) and blended lines complicate the analysis, and can prevent the determination of accurate radial velocities (see C. Proffitt's proceedings).

With the current upgrades on the instruments and facilities, we will soon be able to observe fainter Cepheids with dimmer orbiting companions. The coming GAIA parallaxes will allow us to break the degeneracy between mass and distance, and we will then be able to combine interferometry with single-line velocities to provide dynamical mass measurements of Cepheids.

\begin{acknowledgement} 
\noindent\vskip 0.2cm
\noindent {\em Acknowledgments}: W.G. and G.P. acknowledge support from the BASAL Centro de Astrof\'isica y Tecnologias Afines PFB-06/2007. This research has received funding from the European Research Council under the European Union’s Horizon 2020 research and innovation program (grant No\,695099). PK acknowledges support from the French Agence Nationale de la Recherche (ANR), under grant ANR-15-CE31-0012-01.
\end{acknowledgement}


\begin{thebibliography}{0}

\end{thebibliography}


\begin{thebibliography}{}
%
%




%

\bibitem[{{Gallenne} {et~al.}(2014){Gallenne}, {M{\'e}rand}, {Kervella},
	{Breitfelder}, {Le Bouquin}, {Monnier}, {Gieren}, {Pilecki}, \&
	{Pietrzy{\'n}ski}}]{Gallenne_2014_01_0}
{Gallenne}, A., {M{\'e}rand}, A., {Kervella}, P., {et~al.} 2014, A\&A, 561, L3

\bibitem[{{Gallenne} {et~al.}(2015){Gallenne}, {M{\'e}rand}, {Kervella},
	{Monnier}, {Schaefer}, {Baron}, {Breitfelder}, {Le Bouquin}, {Roettenbacher},
	{Gieren}, {Pietrzy{\'n}ski}, {McAlister}, {ten Brummelaar}, {Sturmann},
	{Sturmann}, {Turner}, {Ridgway}, \& {Kraus}}]{Gallenne_2015_07_0}
{Gallenne}, A., {M{\'e}rand}, A., {Kervella}, P., {et~al.} 2015, A\&A, 579, A68


\bibitem[{{Gallenne} {et~al.}(2013){Gallenne}, {Monnier}, {M{\'e}rand},
	{Kervella}, {Kraus}, {Schaefer}, {Gieren}, {Pietrzy{\'n}ski}, {Szabados},
	{Che}, {Baron}, {Pedretti}, {McAlister}, {ten Brummelaar}, {Sturmann},
	{Sturmann}, {Turner}, {Farrington}, \& {Vargas}}]{Gallenne_2013_04_0}
{Gallenne}, A., {Monnier}, J.~D., {M{\'e}rand}, A., {et~al.} 2013, A\&A, 552,
A21


\end{thebibliography}
%
%

\end{document}